\documentclass[apl,reprint,twocolumn]{revtex4}
\usepackage{amsfonts}
\usepackage{amsmath}
\usepackage{graphicx}
\usepackage{bm}
\usepackage{amssymb}
\usepackage{times}
\usepackage{dcolumn}
 \makeatletter
\def\bib@device#1#2{}
\makeatother

\makeatletter

\begin{document}

\title{Magnetization plateau and incommensurate spin modulation in Ca$_{3}$Co%
$_{2}$O$_{6}$}
\author{Yang Zhao, Shou-Shu Gong, Wei Li, Gang Su$^{{\ast }}$}
\affiliation{College of Physical Sciences, Graduate University of Chinese Academy of
Sciences, P. O. Box 4588, Beijing 100049, People's Republic of China}

\begin{abstract}
The magnetic properties of a trigonal prism unit of the spin-2
frustrated compound Ca$_{3}$Co$_{2}$O$_{6}$ are studied by means of
the density-matrix renormalization group method. A magnetization
plateau at $m_{s}/3$ ($m_{s}$ is the saturation magnetization) with
ferrimagnetic structure is observed. By fitting the experimental
data of
magnetic curve, an estimation of the couplings gives $J_{1}$=$-26.84$K, $%
J_{2}$=$0.39$K, and $J_{3}$=$0.52$K. The local magnetic moments are unveiled
to exhibit an incommensurate sinusoidally modulation along the three chains
of the trigonal prism, which gives a strong theoretical support to the
experimentally observed incommensurate partially disordered
antiferromagnetic state for Ca$_{3}$Co$_{2}$O$_{6}$. The present result suggests that the
modulation indeed originates from the competition of antiferromagnetic and
ferromagnetic couplings.
\end{abstract}

\pacs{75.10.Jm,75.40.Mg,75.50.Ee,75.60.Ej}
\maketitle

Frustrated quantum magnets have been extensively studied owing to their
exotic properties \cite{Frustration}. Among others, the first experimental
realization of ferromagnetic (FM) chains coupled with antiferromagnetic
(AFM) interactions in a triangular plane in the spin frustrated chain
compound Ca$_{3}$Co$_{2}$O$_{6}$ has attracted much attention in the last
decade \cite{Kageyama,Maignan,Hardy}. Ca$_{3}$Co$_{2}$O$_{6}$ consists of
spin chains with alternating face-sharing octahedral (Co I) and trigonal
prismatic (Co II) CoO$_{6}$ polyhedra along the \textit{c} axis, and
arranged in a triangular lattice in the \textit{ab} plane \cite{Kageyama}, as
shown schematically in Fig. 1(a). The Co$^{3+}$ ions on the Co I and Co II
sites have low-spin ($S$=$0$) and high-spin ($S$=$2$) states, respectively\cite%
{NMR}. Within the chains, direct Co-Co overlap leads to a strong FM
intrachain interaction $J_{1}$ (about 25K) \cite{Fresard}. For the
AFM interchain super-superexchange interactions $J_{2}$ and $J_{3}$,
as the overlap of the O $2p$ orbital is very small, $J_{2}$ is
weaker than $J_{3}$ \cite{Fresard}.

A number of techniques (x-ray \cite{Xray1,Xray2}, neutron scattering \cite%
{Neutron}, NMR \cite{NMR}, calorimetry \cite{Calory}, and $\mu $SR \cite%
{Takeshita}, etc.) have been used to study this intriguing compound in
different aspects, such as the nature of the order at different
temperatures, the magnetization process under various conditions, and so on.
It was found that the compound has two transition temperatures $T_{c1}$$%
\simeq $$25$K and $T_{c2}$$\simeq $$7$K, between which the magnetization
curve has a $m_{s}/3$ plateau ($m_{s}$ is the saturation magnetization) \cite%
{Kageyama}. Below $T_{c2}$, the magnetization shows a multistep
structure and the hysteresis, which was suggested in a spin
freezing state \cite{Kageyama,Maignan}, and appears still under debate. Another
intriguing issue is the magnetic state between $T_{c1}$ and $T_{c2}$
in absence of a magnetic field. Owing to the spin frustration, two
possible magnetic structures were proposed: (1) a ferrimagnetic (FI)
structure, with two third of spins up and one third spins down; (2)
a partially disordered antiferromagnetic (PDA) structure, with one
third spins up, one third spins down, and one third spins
incoherent.

\begin{figure}[htbp]
\centering \setlength{\fboxrule}{0.1pt} \setlength{\fboxsep}{0.1mm}
\includegraphics[angle=0,width=0.55\linewidth,
totalheight=45mm]{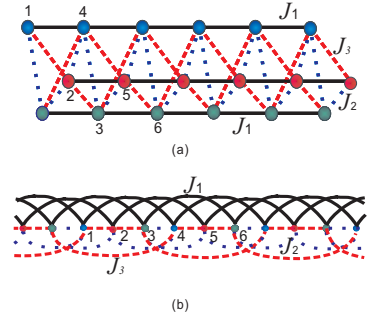} \setlength{\abovecaptionskip}{0mm}
\setlength{\belowcaptionskip}{0mm} \caption{(Color online) (a)
Schematic structure of the trigonal prism unit of
Ca$_{3}$Co$_{2}$O$_{6}$, \emph{J}$_{1}$ (black lines) FM;
\emph{J}$_{2}$ (blue dotted lines) AFM; \emph{J}$_{3}$ (red dashed
lines) AFM. Co II (S=2) sites in different chains are colored
differently. (b) Sketch of the equivalent S=2 spin chain model [Eq.
(\protect\ref{Hamiltonian})].} \label{Structure}
\end{figure}

Recently, the zero-field magnetic state of Ca$_{3}$Co$_{2}$O$_{6}$ between $%
T_{c1}$ and $T_{c2}$ has been shown to be a sinusoidally modulated
incommensurate PDA (iPDA) state with a modulation of very long periodicity ($\sim
$$1000${\AA }) along the chain direction by resonant X-ray scattering
\cite{Xray2,Mazzoli} and neutron diffraction \cite{Agrestini}
measurements. Such a magnetic structure cannot be explained in a
simple two-dimensional (2D) triangular Ising model \cite{Ising}, in
contrast to some previous works. Therefore,
to understand the iPDA state, the trigonal prism unit of Ca$_{3}$Co$_{2}$O$%
_{6}$ including all possible Heisenberg interactions should be considered %
\cite{Agrestini,Chapon}. In the trigonal prism, the AFM interchain couplings
$J_{2}$ and $J_{3}$ follow helical paths and connect the Co II ($S$=$2$) sites
in adjacent chains [Fig. \ref{Structure}(a)]. The competition between the
AFM and FM interactions was argued to be responsible for the modulation
\cite{Agrestini,Chapon}. The emergence
of the iPDA state was suggested to be related to the trigonal prism unit \cite{Chapon}. However, a
direct theoretical analysis of the iPDA state in the trigonal
prism unit of Ca$_{3}$Co$_{2}$O$_{6}$ is still absent.

In this paper, by means of the density-matrix renormalization group (DMRG)
method we shall study the magnetic properties of the $S$=$2$ spin chain model
with the trigonal prism unit under various circumstances, and fit the
experimental result of magnetization curve of Ca$_{3}$Co$_{2}$O$_{6}$ to
estimate the three different couplings $J_{1}$, $J_{2}$ and $J_{3}$.
The origin of the iPDA state in Ca$%
_{3}$Co$_{2}$O$_{6}$ will also be addressed. Our results show that
the system exhibits a $m_{s}/3$ magnetization plateau with a FI
structure, and a very good fitting to the experimental data is
obtained, with the theoretical fitting values in well agreement with
those obtained by other experimental means. The incommensurate
sinusoidally modulated structure with a long period along the chain
direction is observed in the local magnetic moments in the present
model, which may shed light on the origin of the iPDA state in the
compound Ca$_{3}$Co$_{2}$O$_{6}$.

The trigonal prism unit of Ca$_{3}$Co$_{2}$O$_{6}$ [Fig. 1(a)] can
be equivalently expanded to a S=2 Heisenberg spin chain with nearest
neighbor, next, next next, and next next next nearest neighbor
interactions, as illustrated in Fig. \ref{Structure}(b). The
Hamiltonian under interest is given by
\begin{eqnarray}
H &=&\sum_{i=1}^{L/3}[\emph{J}_{1}(\mathbf{S}_{3i-2}\cdot \mathbf{S}_{3i+1}+%
\mathbf{S}_{3i-1}\cdot \mathbf{S}_{3i+2}+\mathbf{S}_{3i}\cdot \mathbf{S}%
_{3i+3})  \notag \\
&&+\emph{J}_{2}(\mathbf{S}_{3i-1}\cdot \mathbf{S}_{3i+1}+\mathbf{S}%
_{3i}\cdot \mathbf{S}_{3i+1}+\mathbf{S}_{3i}\cdot \mathbf{S}_{3i+2})  \notag
\\
&&+\emph{J}_{3}(\mathbf{S}_{3i-2}\cdot \mathbf{S}_{3i-1}+\mathbf{S}%
_{3i-1}\cdot \mathbf{S}_{3i}+\mathbf{S}_{3i-3}\cdot \mathbf{S}_{3i+1})]
\notag \\
&-&h\sum_{j=1}^{L/3}S_{j}^{z},  \label{Hamiltonian}
\end{eqnarray}%
where $\mathbf{S}_{j}$ is the $S$=$2$ spin operator at the \emph{j}-th site, $%
L $ is the total number of spins in the trigonal prism, \emph{J}$_{1}$$<$$0$
(\emph{J}$_{2,3}$$>$$0$) is the FM (AFM) interaction, and $h$ is the
magnetic field applied along the chain direction. For convenience, we take $%
g\mu _{B}$=$1$. During the DMRG \cite{DMRG} iteration, the length of the chain is
taken at least $90$, the optimal states are kept $120$, and the
truncation error is less than 10$^{-5}$.

\begin{figure}[tbp]
\includegraphics[angle=0,width=0.85\linewidth]{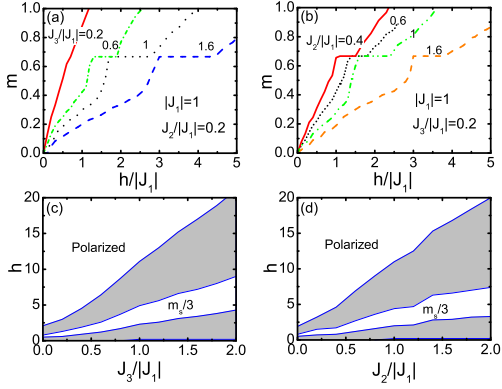}
\caption{(Color online) Magnetization curves of spin-2 trigonal prism chain
with (a) $J_{2}/|J_{1}|$=$0.2$ for different $J_{3}$, and (b) $J_{3}/|J_{1}|$%
=$0.2$ for different $J_{2}$. Magnetic phase diagrams in the $h-J_{2,3}$
plane of the $S$=$2$ trigonal prism chain under interest for (c) $%
J_{2}/|J_{1}|$=$0.2$, and (d) $J_{3}/|J_{1}|$=$0.2$.}
\label{Magnetization}
\end{figure}

Figs. \ref{Magnetization}(a) and (b) show the DMRG
results of the magnetization curve of the $S$=$2$ trigonal prism spin chain system
for different couplings. It can be seen that under proper
couplings, the system has a $m_{s}/3$ magnetization plateau that is
consistent with the necessary condition \cite{OYA} $n(S-m)$=$integer$ for the
appearance of magnetization plateau in Heisenberg quantum antiferromagnets,
where $n$ is the period of ground state, $S$ is the magnitude of spin, and $%
m $ is the magnetization per site. For the present system, $n$=$3$. In
the plateau state, the local magnetic moments are found to have the
same value in each chain of the trigonal prism, arranged in a FI
structure, namely $(-2,2,2)$, in agreement with the experimental
observation \cite{Kageyama}.\textit{\ }The width of the plateau
increases with increasing AFM interactions $J_{2}$ and $J_{3}$, and
decreases with increasing FM coupling $|J_{1}|$, which are
consistent with the behaviors of magnetization plateau observed in
other systems \cite{Gu}. Through massive calculations, the magnetic
phase diagrams in the $h-J_{2}(J_{3})$ plane can be obtained by
observing the critical and saturation magnetic fields of
magnetization curves under different couplings, as depicted in Figs.
\ref{Magnetization}(c) and (d). It appears that
the region for the appearance of $m_{s}/3$ magnetization plateau is
narrow.

\begin{figure}[tbp]
\includegraphics[angle=0,width=0.75\linewidth]{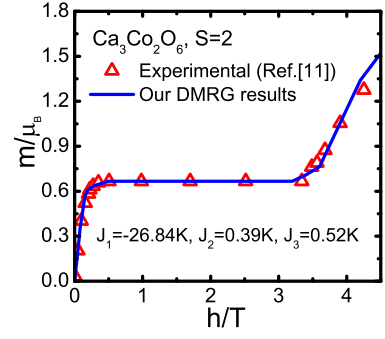}
\caption{(Color online) The experimental data of magnetization curve for Ca$%
_{3}$Co$_{2}$O$_{6}$ at $T$=$10$K from Ref. \onlinecite{Takeshita}
are fitted
with the DMRG results, giving $J_{1}$=$-26.84$K, $J_{2}$=$0.39$K and $J_{3}$=$0.52$%
K. In the calculation, the number of Co sites is chosen as 180.}
\label{Fit}
\end{figure}

The $m_{s}/3$ magnetization plateau of the trigonal prism chain compound Ca$%
_{3}$Co$_{2}$O$_{6}$ has been already observed experimentally %
\cite{Takeshita}. To fit
the experimental data, we make a lot of DMRG calculations \cite{Note}.
The fitting results for the data of $T$=$10$K from Ref. \onlinecite{Takeshita} are
presented in Fig. \ref{Fit}, which shows a good agreement, giving $J_{1}$=$%
-26.84$K, $J_{2}$=$0.39$K, and $J_{3}$=$0.52$K. The so-obtained $J_{1}$ is
very close to $25$K estimated from the experimental study \cite{Kageyama}
and first-principle calculations \cite{Fresard} of Ca$_{3}$Co$_{2}$O$_{6}$, and $%
J_{3}/J_{1}$$\simeq $$-0.019$ is also consistent with the value
$-0.018$ obtained in Ref. \onlinecite{Chapon} within a classical
treatment. Below the plateau, as shown in Fig. \ref{Fit}, the
magnetization curve increases very sharply as indicated by a cusp
observed experimentally \cite{Parras} in the magnetic
susceptibility, which was once regarded as a
feature of FI state of Ca$_{3}$Co$_{2}$O$_{6}$ between $T_{c1}$ and $T_{c2}$%
. Our DMRG results indicate that the iPDA state can also give rise
to a similar behavior below the magnetization plateau.

\begin{figure}[tbp]
\includegraphics[angle=0,width=0.75\linewidth]{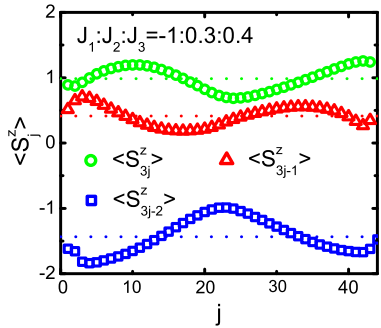}
\caption{(Color online) The spatial dependence of the local magnetic
moments $\langle S^{z}_{j}\rangle$ of the $S$=$2$ trigonal prism spin chain model for
$J_{1}$:$J_{2}$:$J_{3}$=$-1$:$0.3$:$0.4$.} \label{Modulation}
\end{figure}

Now let us look at the incommensurate magnetic modulation observed in Ca$%
_{3} $Co$_{2}$O$_{6}$ through the trigonal prism spin chain
model [Eq. (\ref{Hamiltonian})]. For this purpose, the magnetic moments are
calculated. Figure \ref{Modulation} presents the DMRG result of
local magnetic moment $\langle S_{j}^{z}\rangle $ for
$J_{1}$:$J_{2}$:$J_{3}$=$-1$:$0.3$:$0.4$ as an example. To see clearly the
magnetic modulation, $\langle S_{3j}^{z}\rangle $, $\langle
S_{3j-1}^{z}\rangle $ and $\langle S_{3j-2}^{z}\rangle $, which
correspond to the three chains of the trigonal prism in Fig.
\ref{Structure}(a), are studied, respectively. It is evident that
the spatial dependence of $\langle S_{j}^{z}\rangle $ for the three
chains along the \textit{c} axis exhibits modulations sinusoidally around $%
0.986$ ($\langle S_{3j}^{z}\rangle $), $0.414$ ($\langle
S_{3j-1}^{z}\rangle $), and $-1.4$ ($\langle S_{3j-2}^{z}\rangle $),
indicating that the state is neither FI nor PDA, but an iPDA,
thereby supporting the recent experimental observations
\cite{Xray2,Mazzoli,Agrestini}. During the DMRG calculations, the systems with various chain
lengths are compared. It is found that the expectation value of each local
magnetic moment $\langle S_{j}^{z}\rangle$ is almost independent of
the system size, indicating that the modulation is not a finite-size
effect. In Fig. \ref{Modulation}, we show
the case with $120$ sites (40 unit cells) of the $S$=$2$ spin chain [Eq.
(1)] that gives the period of modulation about $36$ unit cells,
showing that the system can have a magnetic modulation with a long
period. Besides, we have also uncovered the less $|J_{1}|$
($|J_{1}|$$>$$J_{2,3}$), the shorter the modulation period. When $J_{2}$
or $J_{3}$ is set zero and $|J_{1}|$ dominates, the similar
magnetic modulations are also seen. In a general case with $|J_{1}|$$>$$J_{2,3}$%
, the period of modulation is observed to decrease with increasing AFM
interactions $J_{2,3}$ and increase with increasing FM coupling $|J_{1}|$.
These evidences show that the modulation indeed originates from the competition
between the FM and AFM interactions.

It was argued that the magnetic modulation in
Ca$_{3}$Co$_{2}$O$_{6}$ may be caused by both the single-ion anisotropy and the competition of
couplings \cite{Agrestini}. To clarify
this point, we consider the
effect of single-ion anisotropy $-D\sum_{i}(S_{i}^{z})^{2}$
($D$$>$$0$) on the magnetic properties of the spin-2 Heisenberg
complex chain system [Eq. (1)]. The results show that the single-ion
anisotropy weakens the modulation and prolongs the modulation
period. For instance, when $D$=$0.2$, the period exceeds $50$
unit cells. With further increasing $D$, the modulation vanishes and
the local magnetic moments in different chains form a FI structure,
i.e., $(-2,2,2)$ when $D$>$0.4$. When we replace $J_{1}$ terms in Eq. (1) along the three
chains as Ising couplings ($J_{1}$$>$$J_{2,3}$), we find no modulation
for the magnetic moments, implying
that the 2D triangular Ising treatment cannot describe the observed
iPDA state below the plateau.

In summary, the magnetic properties of the $S$=$2$ trigonal prism unit of Ca$%
_{3}$Co$_{2}$O$_{6}$ are studied by means of the DMRG method. The $%
m_{s}/3 $ magnetization plateau with a FI structure is observed, which is consistent with
the experimental observation. The phase diagrams in the $h-J_{2,3}$ plane
are obtained. The experimental data of magnetic curve of Ca$_{3}$Co$_{2}$O$%
_{6}$ are nicely fitted to the DMRG results, giving an estimation of
the interactions $J_{1}$=$-26.84$K, $J_{2}$=$0.39$K, and $J_{3}$=$0.52$K,
in agreement with the previous experimental measurement. The iPDA
state with a long period observed experimentally is supported by the
calculated incommensurate sinusoidally modulated magnetic moments
in the trigonal prism unit of Ca$_{3}$Co$_{2}$O$_{6}$. The effects of
single-ion anisotropy and Ising limit are also
discussed. Our calculation gives a strong support to the proposal
that the iPDA state originates from the frustration due to the
competition between the intrachain FM interaction and the interchain
AFM couplings, and indicates the crucial role of the trigonal prism
structure in understanding the iPDA state of
Ca$_{3}$Co$_{2}$O$_{6}$.

This work is supported in part by the NSFC (Grant Nos. 10625419, 10934008,
90922033), the MOST of China (Grant No. 2006CB601102) and the Chinese
Academy of Sciences.


\begin{thebibliography}{99}
\bibitem{Note} As the present model has a high spin state (S=2) and
contains complex interactions, a direct finite-temperature DMRG
calculation is almost impermissible. Therefore, we perform the DMRG
calculations to fit the magnetization curve at $T$=$10$K.
\end{thebibliography}
\end{document}